\documentclass[12pt, reqno]{amsart}%
\usepackage{amssymb,amsfonts}
\usepackage{subcaption}
\usepackage{graphicx}
\usepackage{xcolor}
\usepackage{stackengine}

\definecolor{green}{rgb}{0.0, 0.8, 0.0}

\usepackage[colorlinks]{hyperref}

\usepackage[margin = 1in]{geometry}

\usepackage[utf8]{inputenc}

\begin{document}

\title{Walking Droplets Through the Lens of Dynamical Systems}

\author{Aminur Rahman\textsuperscript{*,1}}\thanks{\textsuperscript{*}Corresponding Author  (\url{arahman2@uw.edu})}
\thanks{\textsuperscript{1} Department of Applied Mathematics, University of Washington, Seattle, WA 98195}

\author{Denis Blackmore\textsuperscript{2}}\thanks{\textsuperscript{3} Department of Mathematical Sciences, New Jersey Institute of Technology, Newark, NJ 07102}

\begin{abstract}
Over the past decade the study of fluidic droplets bouncing and skipping (or ``walking'') on a vibrating fluid bath has gone from an interesting experiment to a vibrant research field.  The field exhibits challenging fluids problems, potential connections with quantum mechanics, and complex nonlinear dynamics.  We detail advancements in the field of walking droplets through the lens of Dynamical Systems Theory, and outline questions that can be answered using dynamical systems analysis.  The article begins by discussing the history of the fluidic experiments and their resemblance to quantum experiments.  With this physics backdrop, we paint a portrait of the complex nonlinear dynamics present in physical models of various walking droplet systems.  Naturally, these investigations lead to even more questions, and some unsolved problems that are bound to benefit from rigorous Dynamical Systems Analysis are outlined.
\end{abstract}


\maketitle

\section{Introduction}

Dynamical Systems Theory is often used to analyze complex phenomena that arise from the interaction of simpler components.  One problem that comes to mind is the ``$n$-body'' problem.  Solving the $2$-body problem is very simple as shown by Newton \cite{Principia}.  However, as shown by Poincar\'{e} \cite{Poincare1, Poincare2}, adding even a small third body makes the problem quite complex.  Similarly, with walking droplets, even the simplest reduction: a ball bouncing on a table, exhibits chaotic behavior \cite{HolmesTable}.

It has been known since the 1970s that liquid droplets can float and bounce on a vertically sinusoidally accelerated liquid bath \cite{Walker1978}.  Three decades later, Couder and co-workers discovered that increasing the acceleration on the bath causes the droplet to move horizontally in addition to bouncing \cite{CPFB05, CouderFort06}, and thus are dubbed \emph{walking droplets} (or \emph{walkers}).  Incredibly, they had created a classical system where the particle-like object (the droplet) generated waves, which then propelled the droplet.  This was reminiscent of Pilot-wave Theory in Quantum Mechanics, where a wave guides the particle, thereby determining the statistics induced from particle trajectories.  The similarities between walking droplets (or hydrodynamic pilot-waves) and Pilot-wave Theory of Quantum Mechanics, led to extensive research in using walking droplets to conduct Quantum-like experiments \cite{Bush15a, Bush15b}.  While some experiments captured much of the statistics observed in quantum systems, others did not.  As with most complex systems, mathematical models play a pivotal role in analyzing walking droplets and making predictions about their behavior.  The models provide insight into the possible dynamics that have yet to be observed in experiments.  Rigorous dynamical systems analysis can aid experimentalists in fine-tuning the parameters to achieve the desired statistics of particle trajectories \cite{BCGMN}.

There is a rich history of scientific advances through the use of Dynamical Systems Theory.  Sometimes the scientific phenomena inform the theory and other times the theory corrects our understanding of the phenomena.  This is evident in the works of Poincar\'{e} \cite{Poincare1, Poincare2} where he initially had a particular understanding of the $3$-body problem, which worked well for a short period of time, but as he delved further into the analysis he realized that his understanding needed to be modified.  Dynamical Systems Theory has the potential to play a similar role in the walking droplets phenomenon.  This potential has already manifested itself in several analytically rigorous studies, which will be discussed in this review.

The remainder of this article is organized as follows:  In Sec. \ref{Sec: Quantum Mechanics}, some background on relevant quantum mechanical phenomena and interpretations are given.  Section \ref{Sec: Walking Droplets} takes us into the experimental setups that lead to promising analogies with Quantum Mechanics.  Then we discuss the fluid dynamics models in Sec. \ref{Sec: Fluids Models}.  Finally in Sec. \ref{Sec: DS Models} and \ref{Sec: DS Results}, we arrive at the main topic of this review: Dynamical Systems models and interesting results arising from those models.  To conclude, in Sec. \ref{Sec: Unsolved Problems} we outline some unsolved problems that are ripe for rigorous Dynamical Systems analysis.

\section{Quantum Mechanics}\label{Sec: Quantum Mechanics}

Since the nascence of Quantum Mechanics (QM), much of the controversy has been centered around its philosophical implications \cite{BubQuantumBook, BricmontQuantumBook}.  In recent decades it has been shown to be one of the most predictive theories of Physics.  Indeed, it is true that the mathematical foundations of QM are strong.  However, the original controversy of its interpretation lingers.  Some may question the merit in attempting to shed light on the subtleties of the quantum world.  Since the predictions of QM are so precise, why should we not use it to just make measurements and leave the mysterious statistical world underneath undisturbed?  We recall that the ancient Greeks had the ability to measure the location of the planets and stars in the night sky with remarkable precision.  Yet, it was not until Kepler's laws of planetary motion and Newton's laws of gravity that we had both quantitative predictions and a correct qualitative description of the motion of celestial bodies.  Thus, the walking droplets system, due to its similarity to the pilot-wave interpretation \cite{deBroglie, deBroglie1923, deBroglie1956, deBroglie1987, Bohm1952i, Bohm1952ii}, is an intriguing addition to this debate.  Specifically, the walking droplets system is more analgous to de Broglie's ``double-solution'' theory \cite{deBroglie1956} than Bohm's use of a stochastic field \cite{Bohm1952i, Bohm1952ii, Nelson1966JPhys, Nelson1966PhysRev, CouderFort12, Bush15a}.  These and other connections to realist models of quantum dynamics have been explored thoroughly in review articles by Bush \cite{Bush15a, Bush15b} and Bush and Oza \cite{BushOza20_ROPP}.

Unlike classical phenomena, quantum phenomena cannot yet be observed directly.  However, this dilemma is nothing new in science.  A major stumbling block to observing the exact dynamics of quantum particles is in our inability to precisely measure both momentum and position.  One explanation could be that quantum particles are inherently Statistical and do not have definite properties until they are measured.  Similarly, the ancient Greeks developed interpretations based on what they observed in the night sky.  Newton and Huygens, on the other hand, did have access to data from telescopes, but these devices did not completely uncover the realities of planetary motion as our modern day satellites do.  This necessitated analogies to bridge the gap between the available measurements and their unobservable qualitative description.  They realized that a similar phenomenon would be a ball at the end of a string with a radially inward tensile force.  From Newton's writings, it seems that this analogy aided him in developing his theory of gravitation centuries before high resolution observations of the motion of celestial bodies was possible.

\section{Walking Droplets}\label{Sec: Walking Droplets}

\begin{figure}[htbp]
\centering
\includegraphics[width = 0.9\textwidth]{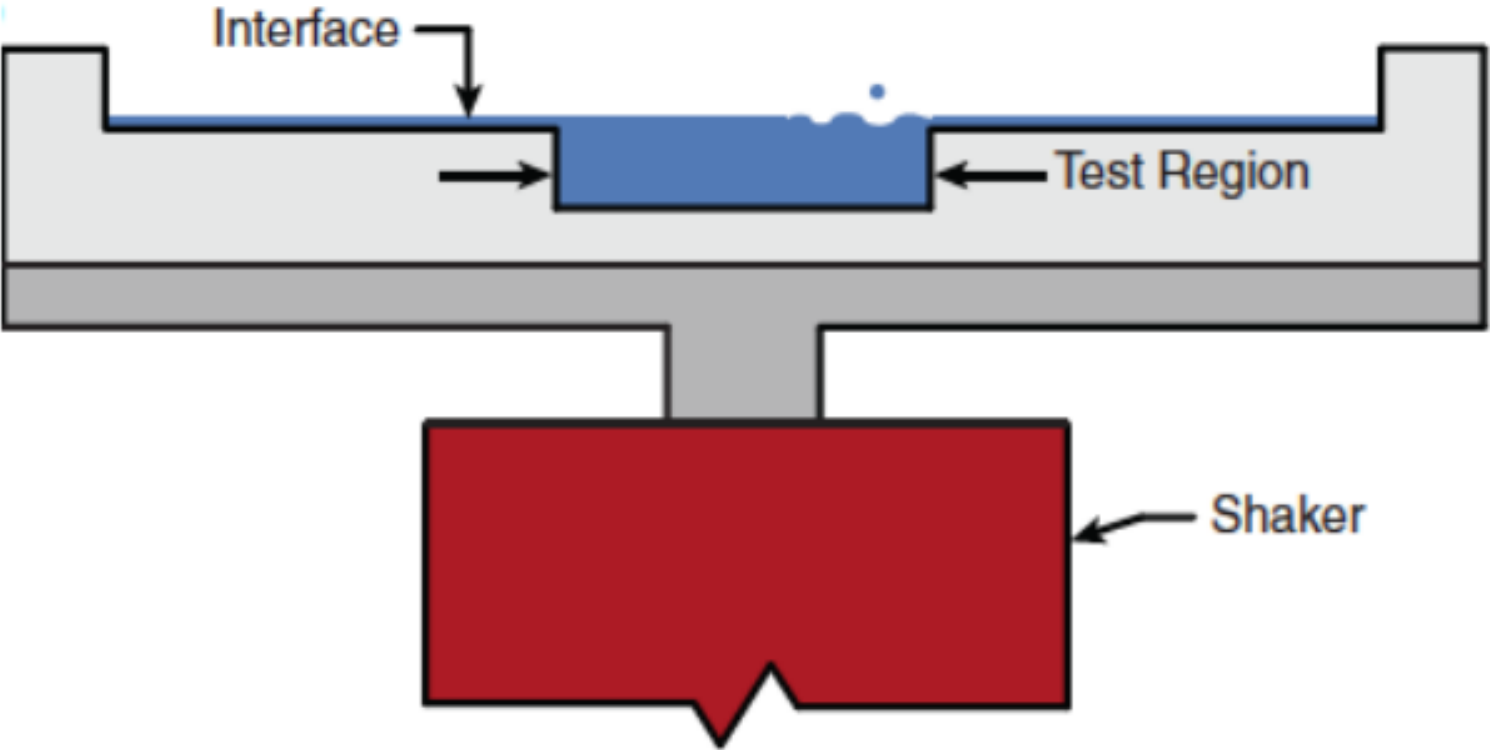}
\caption{Experimental setup to produce walking droplets \cite{HMFCB13}.}
\label{Fig: Experimental Setup}
\end{figure}

In the early days of QM, de Broglie and Bohm championed the interpretation where a pilot-wave governs the motion of a quantum particle \cite{deBroglie, Bohm1952i, Bohm1952ii}.  In recent decades we finally have a directly observable macroscale analog of a pilot wave.  Consider a table that is forced sinusoidally in the vertical direction creating a vibrating fluid bath (Fig. \ref{Fig: Experimental Setup})\cite{HarrisBush13}.  Suppose that the table is forced with an acceleration of
\begin{equation}
    f(t) = \gamma\sin\omega t;\qquad \gamma = A_0\omega^2.
\end{equation}
The surface of the bath is initially flat.  A droplet is dropped onto the bath resulting in excitations to the initially flat wave field.  Also, due to the surface tension of the droplet and the bath, the droplet is propelled upward instead of coalescing with the bath.  A small excitation (small value of $\gamma$) would create a shallow crater in the bath that disappears when the droplet is in the air, and therefore the droplet just bounces.  A larger excitation (large value of $\gamma$) creates a deeper crater that persists even after the droplet returns, and now the droplet can interact with the sides of the crater to be propelled opposite to the slope at that point.  The subsequent impacts of the droplet with the bath contribute to the total wave field.  Now we have a particle (the droplet, which we call a \textit{walker}) being propelled by the waves that it creates; i.e., a pilot wave.

Since the seminal works of Couder and co-workers \cite{CPFB05, CouderFort06, CFGB2005, PBC2006}, there have been many articles on fine-tuning droplet creation and the experimental setup \cite{GMLVD2007, HarrisBush2015, HLB2015}.  In addition, it has since been shown that the air flow in the laboratory contributes to the statistics observed in the system, and therefore more recent experiments enclose the system to isolate the walkers from air \cite{PHFB18, SCB18}.  However, for the purposes of this brief review, we are more interested in documenting the experiments that produce dynamics indicative of nonlinear and chaotic behavior.

Walking was discovered by increasing the acceleration on the fluid bath; i.e., increasing the value of $\gamma$.  There have been extensive studies on the effects of bath acceleration \cite{PCEB2005, PBC2006, MolBush13a, MolBush13b, WMHB13, GVD2007}.  If the bath acceleration is greater than a critical value $\gamma_F$, then standing waves called \emph{Faraday waves} appear on the surface, and $\gamma_F$ is called the \emph{Faraday wave threshold} \cite{Faraday1831}.  Most often experiments are conducted below, but near the Faraday wave threshold, however there are studies of walking above the threshold, which will be discussed in subsequent sections.  From a dynamical systems point of view, the transition from bouncing to walking is particularly interesting due to the bifurcations the system undergoes \cite{MolBush13a, MolBush13b, TLVD13, Blanchette2016}.  While there is plenty of evidence for the observed bifurcations, many have yet to be rigorously proven, which creates a singular opportunity for dynamical systems analysts.  It is common for the fluid dynamics models to be quite complex and not amenable to rigorous dynamical systems analysis so model reduction is often employed.

\begin{figure}[htbp]
    \centering
    \includegraphics[width = 0.9\textwidth]{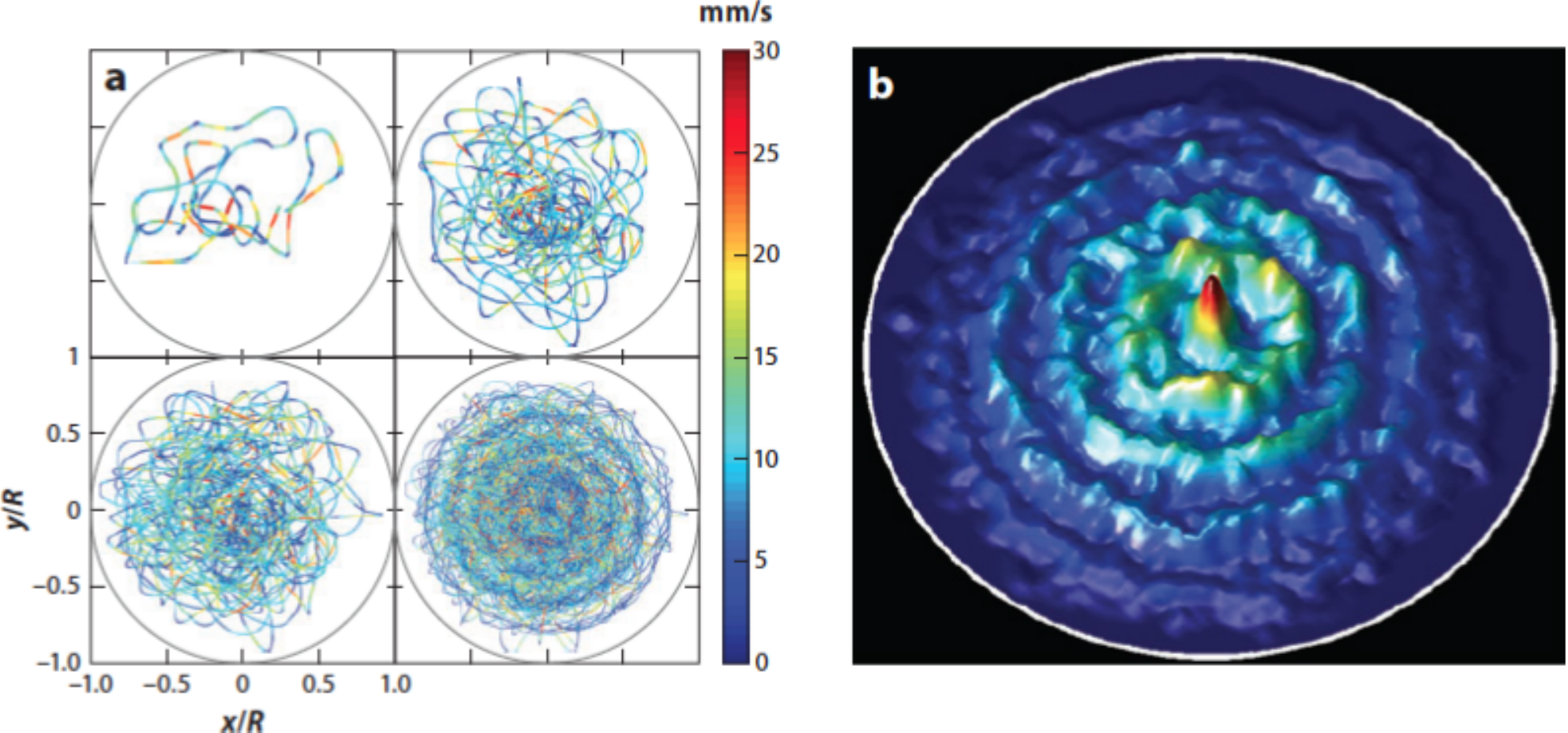}
    \caption{Distribution of walker location preference from the hydrodynamic corral experiments of Harris and Bush \cite{HarrisBush13, HMFCB13}.}
    \label{Fig: Corral Experiment}
\end{figure}

Although there are significant challenges to reproducing quantum statistics for experiments such as the single and double slit phenomena \cite{PHFB18}, corrals and rotating frames produce quite convincing quantum analogs.  Fort \textit{et al.} and Harris and Bush showed that a droplet in a rotating frame will tend towards quantized orbits \cite{FEBMC10, HarrisBush14}.  Due to the Coriolis effect on a mass in a rotating frame being analogous to the Lorenz force on charge in a uniform magnetic field \cite{Bush15a}, further analysis by Oza \textit{et al.} \cite{OHRB14, OWHRB14, ORB18} and Eddi \textit{et al.} \cite{EMPFC12} solidified the analogy of walkers and quantum effects such as \emph{Zeeman splitting}.  Perhaps the most visually compelling example of an analog of a quantum experiment comes from the corral experiments.  In the quantum corral experiment, first conducted by Crommie \textit{et al.} \cite{CLE1993}, a ring of iron atoms are arranged on top of a copper surface causing the surface to form a wave within the boundaries of the ring.  In the hydrodynamic corral experiments of Harris and Bush \cite{HarrisBush13}, they confine a droplet in a circular geometry, and amazingly the droplet produces a position distribution (Fig. \ref{Fig: Corral Experiment}) very similar to the wave produced in the quantum corral.  Later studies by Cristea-Platon, S\'{a}enz, and Bush also produced statistics similar to other quantum experiments \cite{SCB18, CSB18, SCB20}.

Despite the promising experiments in confined geometries, confinement leads to complex boundary interactions, which have proved to be extremely challenging as shown by Pucci \textit{et al.} \cite{PSFB16, PFABB19}.  Simpler quasi -1-dimensional geometries, such as those studied by Filoux \textit{et al.} \cite{FHV15, FHSV17}, are able to confine a droplet without the boundary having a significant impact on the position of the droplet.  These types of confined geometries also naturally allow for dynamical systems models, which are discussed in more detail in Sec. \ref{Sec: DS Models}.

\section{Fluid Dynamics Models}\label{Sec: Fluids Models}

To develop fluid dynamics models of the walker system we need to model the flight of the droplet in the air, the waves on the fluid bath, and the interaction between the droplet and the waves.  Perhaps the simplest form of a particle interacting with a sinusoidally forced surface, which was investigated by Holmes, appears in the form of a ball bouncing on a vibrating table \cite{HolmesTable}.  The system that Holmes studied can be modeled completely using kinematics.  A similar hydrodynamic example was studied by Gilet and Bush where they bounce a droplet on a soap film \cite{GiletBush2009JFM, GiletBush2009PRL}.  These ideas were used by Mol\'{a}\v{c}ek and Bush to develop extremely detailed hydrodynamic models of droplets interacting with a vibrating fluid bath \cite{MolBush13a, MolBush13b}.  While these models captured a significant amount of the physics, they are also computationally expensive to simulate.  Further, the bifurcations and exotic dynamical behavior is obfuscated by the various types of equations coupled to produce the model.  For the purposes of this brief review, we are interested in outlining the dynamical systems results arising from the walking droplets system.

Much of the present bifurcation and chaotic behavior for bouncing can be inferred from the studies of Holmes \cite{HolmesTable} and Gilet and Bush \cite{GiletBush2009JFM, GiletBush2009PRL}.  When the droplet goes from the bouncing to the walking regime, it undergoes period-doubling bifurcations, which were first observed in the experimental study of Mol\'{a}\v{c}ek and Bush \cite{MolBush13a, MolBush13b}.  In order to truly analyze the dynamics, we must make some simplifying assumptions.  Oza \textit{et al.} posited that the excitation from an impact travels fast enough through the bath before the subsequent impact that the droplet feels the same effect as it would if the excitation was instaneously a Bessel function \cite{ORB13}.  This was dubbed the \emph{stroboscopic model} since we are not interested in what happens between bounces.

For the stroboscopic model of Oza \textit{et al.} \cite{ORB13}, let us consider a case when the droplet creates a single impact at $x_0$ and then is taken away. In the following derivations we will assume the variables have been nondimensionalized, and all parameters have been reduced as much as possible.  This single impact on the flat bath creates an excitation that takes the shape of a Bessel function of the first kind centered at and symmetric about $x_0$, $J_0(|x-x_0|)$ \cite{ORB13, ESMFRC11}.  Since the system has natural damping due to the viscosity of the fluid, the excitation will decay exponentially over time, $t$, and hence the excitation becomes 
\begin{equation}
    y = h_0(x,t) = J_0(\omega|x-x_0|)e^{-t}.
    \label{Eq: Oza Wavefield}
\end{equation}
In reality, the droplet falls back down after some time.  Then the bath propels the droplet in a direction opposite to the gradient of the wavefield.  The impact also contributes a new excitation to the wavefield, which is added to the old one to produce the new wavefield.  All of the excitations are summed up to produce the total wavefield, $H(x,t)$.  After each impact it is assumed that that droplet behaves like a projectile, and therefore the position, $x$, of the droplet is modeled as
\begin{equation}
    m\ddot{x} + D\dot{x} = -mg\nabla H(x,t)
    \label{Eq: Oza Model}
\end{equation}
where D is the drag coefficient on the droplet \cite{ORB13}.  For a much more detailed derivation, the reader is referred to the works of Oza \textit{et al.} \cite{ORB13} and Mol\'{a}\v{c}ek and Bush \cite{MolBush13a, MolBush13b}.

While the stroboscopic model provides a natural transition to dynamical systems models, other types of models have numerically shown to exhibit intersting dynamical behavior.  For example, the works of Milewski \textit{et al.} \cite{MGNB15}, Durey \textit{et al.} \cite{DureyMilewski2017, DMW20}, Galeano-Rios \textit{et al.} \cite{GMV17}, Faria \cite{Faria2017}, and others employ a quasi-potential wave model with various forms of droplet motion.  This was recently extended to the experiments of Valani \textit{et al.} on superwalkers \cite{VSS19} by Galeano-Rios \textit{et al.} \cite{GMV19} where the bath is forced at two driving frequencies.  Although the original stroboscopic model is designed for the high memory regime, and indeed most of the quantum-like statistics arise within this regime, interesting behavior is also observed in the low memory regime.  In fact, some of the first such models only considered a single previous impact \cite{PBC2006}, similar to the assumption made earlier in this Section.  In the low memory regime there are also electrodynamics analogs such as a boost to the walker's effective mass provided by the wavefield as shown by Bush \textit{et al.} \cite{BOM14} and Labousse and Perrard \cite{LaboussePerrard2014}.

One major challenge in modeling walking droplets are boundary effects.  These are still not well understood, but there has been significant effort from both the experimental, which was discussed in the previous section, and theoretical sides, in resolving the dynamics at boundaries.  If specular reflection is assumed, then the droplet dynamics is similar to that of billiards \cite{Shirokoff13}: an observation that may have been useful.  However, it was shown that this is not the case \cite{PSFB16}, and therefore boundary interactions need to be studied in more detail.

\section{Dynamical Systems Models}\label{Sec: DS Models}

In this section we shall present a concise but rather complete description of
most of the leading mathematical models of walking single droplet
phenomena, with a focus on those that promise the most fertile ground for
further rigorous dynamical systems analysis. We begin with continuous
dynamical systems (differential equations) \ models, which is only natural and
fitting, since they represent the first realistic, hydrodynamic science based
efforts to describe and predict the motion of walking droplets. Then, we turn
our attention to discrete dynamical systems (difference equations) models,
which for the most part have been derived formally or in an informed intuitive
manner from continuous formulations.

\subsection{Continuous dynamical models}

Starting with pioneering efforts based on the fundamentals of physics
complemented by observation of experiments, in Proti\`{e}re \emph{et al}.
\cite{PBC2006}, Eddi \emph{et al}. \cite{ESMFRC11} and Fort \emph{et al}.
\cite{FEBMC10}, dynamical systems based modeling of walking droplet (pilot-wave)
phenomena has come a long way. More precisely, building upon the pioneering
work with the aid of several realistic simplifying assumptions,
Mol\'{a}\v{c}ek \& Bush \cite{MolBush13a,MolBush13b} and Oza \emph{et al}. \cite{ORB13} developed a
delay dynamical system model of the following form for the horizontal and
vertical motion of a single walking droplet:%
\begin{equation*}
m\ddot{\boldsymbol{r}}+D\boldsymbol{\dot{r}}=-mg\nabla h(\boldsymbol{r}%
,\boldsymbol{r}_{(t)},t)-F(\boldsymbol{r}),\;\ddot{z}+m\left[  g+\gamma
(t)\right]  =Z\left(  \bar{z},\dot{\bar{z}};k,b\right),
\end{equation*}
where $\dot{}:=d/dt$, $\boldsymbol{r}$ is a point in the region $\Omega
\subset\mathbb{R}^{q}$, $q=1,2,$, representing the
oscillating floor of the container of the liquid bath, $z\in\mathbb{R}$ is the
vertical position of the bottom of the droplet with respect to the undeformed
initial surface of the bath, $m$ is the mass of the droplet, $g$ is the
gravitational acceleration, and $\gamma(t)$ is the vertical acceleration of the
vibrating bath.  Here $F(\boldsymbol{r})$ is a possible horizontal force such as
occurs in a rotating corral, $\bar{z}:=z-$ $h(\boldsymbol{r},\boldsymbol{r}%
_{(t)},t)$, $\nabla$ is the gradient operator with respect to $\boldsymbol{r}%
$, $k$ and $b$ are, respectively, spring and damping constant parameters,
$Z\left(  \bar{z},\dot{\bar{z}};k,b\right)  $ is a smooth approximation of
$H\left(  -\bar{z}\right)  \max\left(  -k\bar{z}-b\dot{\bar{z}},0\right)  $,
with $H$ the Heaviside step function.  Also, $h(\boldsymbol{r},\boldsymbol{r}%
_{(t)},t)$ is the wave field height of the oscillating (free) surface of the
bath, which is a function of $\boldsymbol{r},t$ and the delay $\boldsymbol{r}%
_{(t)}$ which denotes the function $\boldsymbol{r}:[0,\infty)\rightarrow
\mathbb{R}^{q}$ restricted to $[0,t]$, which we expect to be smooth.\emph{\ }%
It is precisely in $h$ that the delay occurs, which is also referred to as
\emph{path-memory}.

The above system involves delay differential equations, which lead to
infinite-dimensional dynamical systems that tend to be difficult to analyze
(see \cite{AHA,JRob,Temam}). To be a bit more specific, the phase space vector
and infinite-dimensional dynamical system, respectively, take the form
$u:=\left(  \boldsymbol{r},\boldsymbol{\dot{r}},z,\dot{z},h\right)  ,\;\dot
{u}=\mathcal{F}(u)$ on the infinite-dimensional manifold $\mathcal{M}%
:=\mathbb{R}^{2(l+1)}\times C^{1}\left(  [0,\infty),\mathbb{R}^{l}\right)  $,
where $C^{1}\left(  [0,\infty),\mathbb{R}^{l}\right)  $ denotes the
continuously differentiable functions mapping the interval $[0,\infty)$ into
$\mathbb{R}^{l}$.

Fortunately, a justifiable additional assumption enables the delay
differential equations to be recast as the following integro-differential
equation (IDE) system:%
\begin{subequations}%
\begin{align}
m\ddot{\boldsymbol{r}}+D\boldsymbol{\dot{r}}  &  =-mg\nabla h(\boldsymbol{r}%
,t)-F(\boldsymbol{r,\dot{r}})\label{e5.1a}\\
\ddot{z}  &  =-m\left[  g+\gamma(t)\right]  +Z\left(  \bar{z},\dot{\bar{z}%
};k,b\right)\label{e5.1b}
\end{align}
\label{e5.1}%
\end{subequations}
where
\begin{equation}
h(\boldsymbol{r},t):=\frac{A}{T_{F}}{\displaystyle\int\nolimits_{-\infty}^{t}%
}J_{0}\left(  k_{F}\left\vert \boldsymbol{r}-\boldsymbol{r}(\tau)\right\vert
\right)  e^{-(t-\tau)/T_{F}Me}d\tau, \label{e5.2}%
\end{equation}
with the understanding that $\boldsymbol{r}(\tau):=\boldsymbol{r}(0)$ for
$\tau\leq0$. In the above, $A$ is the initial amplitude of the generated
waves, $T_{F}$ is the \emph{Faraday time} between bounces, $J_{0}$ is the
zeroth order Bessel function of the first kind, $k_{F}:=2\pi/\lambda_{F}$ is
the wave number corresponding to the least stable Faraday wave with wavelength
$\lambda_{F}$ and $Me$ is the \emph{memory} parameter that determines the
damping speed of prior bounce generated waves. The first equation of
\eqref{e5.1} is usually referred to as the \emph{trajectory equation}, which
determines the \emph{horizontal dynamics}, while the second equation is said
to describe the\emph{ vertical dynamics}. It should be noted that the vertical
dynamics \eqref{e5.1b} can be modeled in several ways, often as a linear or nonlinear
spring \cite{GiletBush2009JFM, GiletBush2009PRL, MolBush13b, GMV17}.

The system \eqref{e5.1} with \eqref{e5.2} still represents an
infinite-dimensional dynamical system (in fact, typically $\aleph_{0}%
$-dimensional) of the form
\begin{equation}
\dot{u}=\mathfrak{F}(u;\kappa) \label{e5.3}%
\end{equation}
with $u=\left(  \boldsymbol{r},\boldsymbol{\dot{r}},z,\dot{z},h\right)  $ on
the infinite-dimensional manifold $\mathcal{M}:=\mathbb{R}^{2(l+1)}\times
C^{1}\left(  [0,\infty),\mathbb{R}^{l}\right)  $ and $\kappa$ represents the
vector of physical parameters associated with the system, but it is easier to
study via numerical simulation and appears to be more amenable to extensive
dynamical systems analysis.

Penultimately, we mention the quasi-potential formulation of \ Milewski
\emph{et al}.\cite{MGNB15} \ not only because it represents a more complete
incorporation of fundamental fluid dynamics principles than most other extant
models, but it also can be reduced to a form with a corresponding discrete
dynamical system, which we shall briefly describe in the next subsection. The
model is for an infinite domain and the wave dynamics can be conveniently
expressed in cylindrical coordinates $(r,\theta,z)$, where one is able, by
making very reasonable assumptions, to describe the model in terms of a
velocity potential \ $\phi(r,\theta,z,t)$ and a perturbation $\eta
(r,\theta,t))$ satisfying the following evolution equations and auxiliary conditions:%
\begin{subequations}%
\begin{align}
\phi_{t}  &  =-g_{L}(t)+2\nu\Delta_{H}\phi+\left(  \sigma/\rho\right)
\Delta_{H}\eta-P(r,\theta,t),\;z=0\\
\eta_{t}  &  =\phi_{t}+2\nu\Delta_{H}\eta,\;z=0,
\end{align}
\label{e5.4}%
\end{subequations}
and%
\begin{equation}
\Delta\phi=0,\text{ for }z\leq0,\text{ }\nabla\phi\rightarrow0\text{ and }%
\eta\rightarrow0\text{ as }r^{2}+z^{2}\rightarrow\infty, \label{e5.5}%
\end{equation}
where $g_{L}$ is the effective gravity in the vibrating frame, $\sigma$ is the
tension, $\rho$ is the constant density and $\nu$ the kinematic viscosity,
respectively, of the incompressible fluid bath, $\Delta_{H}\phi:=r^{-1}%
(r\phi_{r})_{r}+r^{-2}\phi_{\theta\theta}$ is the horizontal Laplacian,
$\Delta\phi:=r^{-1}(r\phi_{r})_{r}+r^{-2}\phi_{\theta\theta}+\phi_{zz}$ is the
Laplacian, and $P$ is the droplet impact pressure on the bath surface. These
equations are then combined with the droplet dynamics, which may be taken as
essentially the same as in second equation of \eqref{e5.1}, to obtain an
infinite-dimensional dynamical system of the form
\begin{equation}
\dot{w}=\mathfrak{M}(w;\kappa), \label{e5.6}%
\end{equation}
for $w:=\left(  \phi,\eta,\boldsymbol{r},\boldsymbol{\dot{r}},z,\dot
{z}\right)  $ on an infinite-dimensional manifold of the form $\mathcal{M}%
_{M}:=\mathcal{M}_{\phi}\times\mathcal{M}_{\eta}\times\mathbb{R}^{6}$, where
$\mathcal{M}_{\phi}$ comprises the functions $\phi$ satisfying \eqref{e5.5}
for all $t\geq0$ and $\mathcal{M}_{\eta}$ consists of all $C^{2}$ functions
$\eta$ satisfying \eqref{e5.5} for all $t\geq0$. As one might expect, the
extra attention to physical details in the dynamical system \eqref{e5.6} makes
it more difficult to simulate and analyze than \eqref{e5.3}. A reduced form of
the system that is solvable numerically at a much smaller computational cost
was developed by Durey and Milewski \cite{DureyMilewski2017}, and extended by Faria
\cite{Faria2017}.  Similarly, the stroboscopic model was reduced to a Lorenz-like
dynamical system in an investigation by Durey \cite{Durey2020}.  Finally, we call attention 
to the interesting quasi-potential model of
Galeano-Rios \emph{et al}. \cite{GMV17}, which can also be extended to model superwalkers
\cite{GMV19}.

\medskip

\noindent\textbf{Remark 1}. It is interesting to observe that all the
dynamical systems models considered in this section (and most of the others not
treated here) comprise two coupled components: a (horizontal) trajectory
equation; and a (vertical) droplet dynamics equation. Moreover, the coupling
is essentially one-way, inasmuch as it is possible to solve for the trajectory
independently of the droplet motion owing to the fact that this entails the
calculation of the wave field on the surface of the liquid bath, but
determination of the droplet dynamics requires the solution of the trajectory
equation because it involves the wave field. This probably explains why
droplet dynamics often played a secondary role early on in hydrodynamic pilot-wave research.  In recent years, droplet dynamics has been receiving increasingly more attention necessitating the use of Dynamical Systems Theory.

\subsection{Discrete dynamical system models}

Continuous dynamical systems such as those treated above can often, via a
variety of methods that include simplifying assumptions, reduction techniques,
and experimentally-based inferences, be used to obtain corresponding discrete
dynamical systems models, which are invariably easier to simulate and analyze
than the systems from which they are derived. Now given any continuous
dynamical system, finite or infinite dimensional, of the form $\dot{u}=G(u),$
one can always obtain a discrete dynamical approximation via numerical
integration, where the accuracy depends on the choice of the method and the
increments chosen for the variables. Such methods include complete finite
difference or finite element approaches and semidiscrete schemes in which the
space variables are handled via finite difference, finite element, spectral, or
pseudo-spectral methods that produce a system of ordinary differential
equations that are resolved using one-step techniques such as Runge--Kutta or
multistep methods such as those of the Adams--Moulton predictor-corrector
type, or some variation of these (see \cite{PTVF}).

An example of a variant of the numerical integration schemes can be
illustrated in outline using the system \eqref{e5.1}. As mentioned in Remark
1, it essentially suffices to show how the first (trajectory) IDE can be
integrated numerically. First, we introduce the variable $\boldsymbol{w}%
:=(\boldsymbol{r},\boldsymbol{v}):=(\boldsymbol{r},\boldsymbol{\dot{r}})$,
whereupon the trajectory equation can be recast as
\begin{equation}
\boldsymbol{\dot{w}=\Phi(w},t;\kappa\boldsymbol{)}:=\left(  \boldsymbol{v}%
,-\frac{1}{m}(D\boldsymbol{v}+F(\boldsymbol{r})-g\nabla h(\boldsymbol{r}%
,t))\right)  . \label{e5.7}%
\end{equation}
This equation can then be converted to a finite difference approximation using
a standard one-step integrator such as the Runge-Kutta method, with the only
wrinkle being that the integral has to be approximated by a standard numerical
scheme. For another example - one of standard semidiscrete type - consider the
system of partial differential equations \eqref{e5.4}, which for convenience
we consider on a bounded domain with boundary conditions and harmonic
constraint as in \eqref{e5.5}. Then the spatial requirements, including the
harmonic constraint can be approximated by finite differences, after which one
has a system of ODEs for the approximations, which could be solved by an
Adams--Moulton predictor-corrector scheme.

A specific example similar to approaches sketched above was developed and
studied by Shirokoff \cite{Shirokoff13}. He essentially begins with the trajectory
equation of \eqref{e5.1}, makes some plausible simplifying assumptions
including periodic droplet impacts of period $T$, and then integrates over a
period to obtain the following $\aleph_{0}$-dimensional difference equation
system for the position and velocity at the $(n+1)^{\text{th}}$ impact:%
\begin{equation}
\left(  \boldsymbol{y}_{n+1},\boldsymbol{v}_{n+1}\right)  =\left(
\boldsymbol{y}_{n}+\boldsymbol{v}_{n},v_{n}+R\left(  v_{n},y_{n}%
,y_{n-1},\ldots y_{1};\kappa\right)  \right)  , \label{e5.8}%
\end{equation}
where $\boldsymbol{y}$ is the position of the droplet in the domain
$\Omega\subset\mathbb{R}^{2}$, $\boldsymbol{v}$ is the velocity
$\boldsymbol{\dot{y}}$, $\kappa$ is, as usual, the physical parameter vector
and $R$ is a $C^{1}$ map. The infinite dimensionality of the above system
makes even it difficult to simulate, so Shirokoff makes a further short memory
assumption: namely, that the function $R$ above does not depend on $y_{k}$ for
$k<n-1$. Consequently, the discrete dynamical system \eqref{e5.7} can be
reduced to the following 8-dimensional model for $\boldsymbol{u}_{n}:=\left(
\boldsymbol{y}_{n-1},\boldsymbol{y}_{n},\boldsymbol{v}_{n}\right)  $:%
\begin{equation}
\boldsymbol{u}_{n+1}=S\left(  \boldsymbol{u}_{n};\kappa\right)  :=\left(
\boldsymbol{y}_{n},\boldsymbol{y}_{n}+\boldsymbol{v}_{n},v_{n}+R\left(
v_{n},y_{n},y_{n-1},\ldots y_{1};\kappa\right)  \right)  , \label{e5.9}%
\end{equation}
which is much easier to analyze and simulate than \eqref{e5.8}.

Before describing what appears to be the simplest discrete dynamical system
capable of exhibiting a good deal of the dynamics observed in experiments and
simulations of more complicated mathematical models, we shall briefly mention
a discrete dynamical model obtained from the system \eqref{e5.6}. Durey and
Milewski \cite{DureyMilewski2017} show how \eqref{e5.6} can be reduced so that it can be
formulated as a rather simple homogeneous system with jump conditions. This
reformulation in essence represents the droplet-bath interactions as
instantaneous, thereby enabling the system to be recast as a discrete
dynamical system of the form
\begin{equation}
\boldsymbol{\xi}_{n+1}=\mathcal{M}\left(  \boldsymbol{\xi}_{n},\kappa\right)
, \label{e5.10}%
\end{equation}
where the $\boldsymbol{\xi}_{n}$ are $\aleph_{0}$-dimensional vectors
containing droplet position information and wave field characterizations of
the wave field of the the oscillating bath surface in term of coefficients of
the corresponding Fourier--Bessel expansions. System \eqref{e5.10} is indeed
complicated, but can be approximated by finite-dimensional systems by
truncating the Fourier--Bessel expansions of the wave fields.

For our final discrete dynamical system model for walking droplets, we
consider the ``toy'' models of Gilet \cite{Gilet14,Gilet16}, which he described as the
simplest model that he thought capable of capturing a good deal of
experimentally observed, confined wave-particle coupling dynamics. As it turns
out, Gilet's toy models do indeed exhibit quite a few of the dynamical regimes
- even some of the most complex - observed in experiments and obtained from
simulations of more physically complete dynamical systems representations of
walking droplet phenomena.

\begin{figure}[htbp]
\centering
\includegraphics[width = 0.9\textwidth]{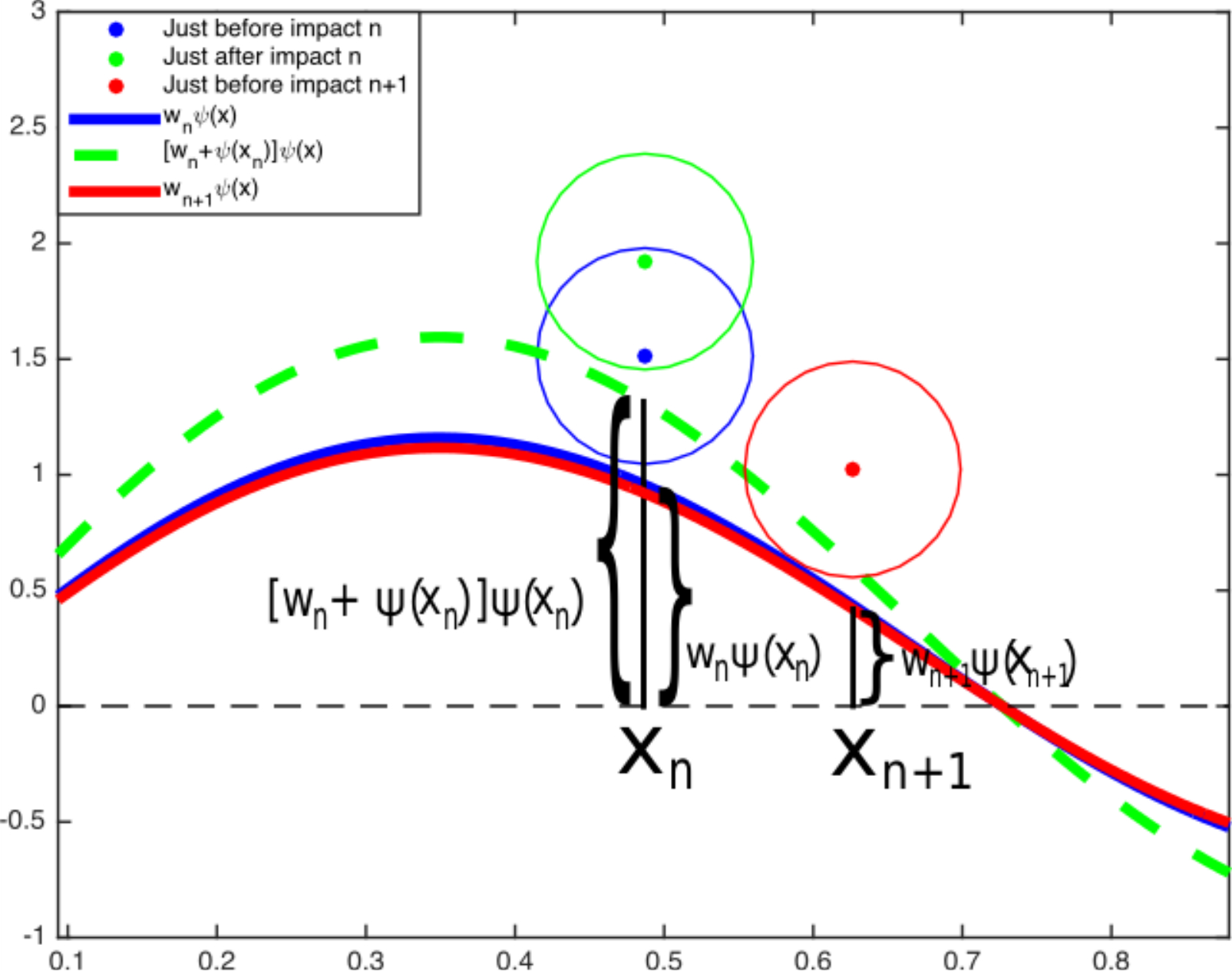}
\caption{Illustration of the model \eqref{e5.11} \cite{RahmanBlackmore16}.}\label{Fig: GiletModel}
\end{figure}

We begin with Gilet's 1-dimensional bath model, where the domain is a finite
interval, say $\Omega:=[\alpha,\beta]$, which can be considered as
$[-\pi/2,3\pi/2]$ by scaling. The domain can then be specified by the
eigenfunctions of the 1-dimensional Laplace operator $\Delta_{1}:=d^{2}%
/dx^{2}$ with Dirichlet or Neumann boundary conditions. For example, in the
latter case we note that the basis of eigenfunctions is $E_{N}(\Omega
)=\{\sin(2n+1)x:n\in\mathbb{Z},n\geq0\},$ where $\mathbb{Z}$ is the integers,
and in what follows we define an \emph{eigenmode }$\Psi$ for this and higher
dimensional models to be a $C^{1}$(Fourier) expansion in the basis of
Laplacian eigenfunctions for the domain $\Omega$. Gilet's 1-dimensional bath
model is a (vector) difference equation of the form
\begin{subequations}%
\begin{align}
x_{n+1}  &  =x_{n}-Cz_{n}\Psi^{\prime}(x_{n})\\
z_{n+1}  &  =\mu\left[  z_{n}+\Psi(x_{n})\right]  ,
\end{align}
\label{e5.11}%
\end{subequations}
where $x_{n}$ is the position of the droplet along the bath, $C\in\lbrack0,1]$
represents the wave-particle coupling, $z_{n}$ is the wave field amplitude on
the bath surface just prior to an impact, so it can be considered as defining
the vertical dynamics of the droplet, $\Psi$ and $\Psi^{\prime}$ are,
respectively, the (assumed) single eigenmode and its spatial derivative and
$\mu\in\lbrack0,1]$ is the path-memory related damping factor. An illustration
of the model is shown in Fig. \ref{Fig: GiletModel}.
Owing to its
form and the periodicity of the eigenmode, one can conveniently extend the
system \eqref{e5.11} to the whole $x,z$-plane, so the dynamics can be
represented by the iterates of the map $G_{1}:\mathbb{R}^{2}\rightarrow
\mathbb{R}^{2}$defined as
\begin{equation}
G_{1}(x,z):=\left(  x-Cz\Psi^{\prime}(x),\mu\left[  z+\Psi(x)\right]  \right)
. \label{e5.12}%
\end{equation}

Not surprisingly, Gilet's 2-dimensional bath model for $\Omega\subset
\mathbb{R}^{2}$, assuming a single eigenmode $\Psi$ with a $C^{1}$ expansions
in terms of the Dirichlet or Neumann eigenfunctions on $\Omega$ of
$\Delta:=\partial^{2}/\partial x^{2}+\partial^{2}/\partial y^{2},$ is a map
$G_{2}:\mathbb{R}^{3}\rightarrow\mathbb{R}^{3}$ defined as
\begin{equation}
G_{2}(x,y,z):=\left(  x-Cz\Psi_{x}(x,y),y-Cz\Psi_{y}(x,y),\mu\left[
z+\Psi(x,y)\right]  \right)  , \label{e5.13}%
\end{equation}
where the eigenmodes $\Psi$ would be represented in Fourier--Bessel series and
Fourier--Mathieu series for circular and elliptic corrals, respectively.
Clearly, Gilet's models are easy to simulate and they turn out to be
comparatively amenable to rigorous dynamical systems analysis.

Another type of discrete dynamical model, developed by Rahman \cite{Rahman18}, models multiple walkers with a bath acceleration below the Faraday wave threshold (describing the experiments of Filoux \textit{et al.} \cite{FHV15}) and a single walker with the bath acceleration above the Faraday wave threshold.  For the single walker model, pitchfork and period doubling bifurcations, and chaos is observed and rigorously proved via Dynamical Systems Analysis.  In this model, it is assumed that the velocity receives a ``kick'' at each impact, and therefore is similar to the Standard map \cite{Chirikov1, Chirikov2, Ott}.  The velocity becomes
\begin{equation}
v_{n+1} = f(v_n,\theta_n) = C\left[v_n + K\sin(\omega(\theta_n - \theta_{n-1}))e^{-\nu (\theta_n - \theta_{n-1})^2}\right],
\label{Eq: SingleVelocity}
\end{equation}
where $v_n$ and $\theta_n$ are the velocity and positions of the single droplet after the $n\textsuperscript{th}$ impact.  The nondimensional wavenumber is $\omega = 2\pi(R_{\text{in}} + D/2)/\lambda_F$, where $\lambda_F$ is the Faraday wavelength and $R_{\text{in}} + D/2$ are the averages of the inner and outer radii.  The spatial damping of the kick strength is represented by the exponential term with a damping parameter $\nu\in \mathbb{R}^+$, which implicitly depends on $\lambda_F$ and the memory of the system.  Then, just as with the Standard map, if $\theta_{n+1} = \theta_n + v_{n+1}$,
\begin{equation}
f(v) := C\left[v + K\sin(\omega v)e^{-\nu v^2}\right].
\label{Eq: SingleMap}
\end{equation}
The new 1-dimensional map preserved significant qualitative features of walkers on an annulus, and was well-suited for rigorous Dynamical Systems analysis.

\section{Dynamical Systems Results and Conjectures}\label{Sec: DS Results}

Now we shall encapsulate some of the more important dynamical systems related
results obtained from studies of the mathematical models described in the
preceding section, followed by some conjectures based upon the knowledge
distilled from the investigations. The results shall be delineated in the
context of modern dynamical systems theory\cite{AP,New,PdM,CRob,JRob,Smale,Temam,Wig}.
As we shall see, the complexities
of most of the extant models for walking droplet dynamics have led to a
preponderance of results being of the experimental or numerical simulation variety.

\subsection{Results obtained from investigation of dynamical systems models}

Using Section 5 as a guide, we begin with the continuous and end with the
discrete dynamical systems models. Although the quantum analogs of walking
droplet phenomena are fascinating, our focus shall be almost exclusively the
dynamical systems related outcomes.

\subsubsection{Outcomes for continuous dynamical systems models}

The early investigations of walking droplet phenomena produced a number of
interesting dynamical systems related observations and conclusions. For
example, Proti\`{e}re \emph{et al}. \cite{PBC2006} reported symmetry breaking,
pitchfork, and period-doubling bifurcations based upon experimental
observations and approximate solution of a relatively simple model for the
trajectory dynamics. Several years later, employing experiments and a
phenomenological model, Eddi \emph{et al}. \cite{ESMFRC11} also observed
symmetry breaking and period-doubling in single droplet period trajectory
dynamics as well as travelling wave dynamics and indications of transition to
chaotic regimes. Fort \emph{et al}. \cite{FEBMC10} investigated the motion of a
single walking droplet in a rotating circular corral both experimentally and
using a trajectory equation that might be considered a simplified version of
that in the system \eqref{e5.1}. Experiments and numerical solution of their
trajectory equation showed the existence of stable periodic (circular)
trajectories occurring in discrete sets in quantum-like fashion.  Orbital
quantization and the emergence of quantum-like
statistics were later considered by Perrard \textit{et al.} \cite{PLMFC14}, Harris and Bush
\cite{HarrisBush14}, Bush \cite{Bush15a}, and Gilet \cite{Gilet16}.

More recent studies of continuous evolution mathematical models of walking
droplet phenomena, which along with much of the material in this paper, have
been very adroitly summarized in such papers as Bush \cite{Bush15a,Bush15b, BushOza20_ROPP} and
Turton \emph{et al}. \cite{TCB}, have uncovered many more results amounting to
an extensive menagerie of dynamic possibilities. Many of the more contemporary
investigations have been inspired by the seminal work of Mol\'{a}\v{c}ek and
Bush \cite{MolBush13a,MolBush13b} and Oza \emph{et al}. \cite{ORB13}. The results of the
investigations in \cite{MolBush13a,MolBush13b}, which comprise both experimental and
theoretical components, include a dynamical model with a trajectory equation
that is essentially a simplified precursor of that in \eqref{e5.1}, together,
ultimately, with a logarithmic spring vertical dynamics equation. Numerical
integration of their model showed the existence of pitchfork bifurcations for
a sufficiently large forcing acceleration $\gamma$ (or memory $Me$) followed
by period-doubling cascades leading to chaos as the parameter increased;
outcomes that were in good agreement with experimental results. Oza \emph{et
al}. \cite{ORB13} devised the trajectory equation in \eqref{e5.1}, which has
proven to be one of the most reliable qualitative and quantitative predictors
of horizontal walking droplet dynamics. Using analysis and some numerics on
their equation, they demonstrated the existence of a supercritical pitchfork
bifurcation for sufficiently large $\gamma$ (or $Me$) that generates a
rectilinear trajectory. Moreover, they showed that this trajectory tends to
destabilize transversely with increasing $Me$, suggesting that analogous
destabilization of more complicated orbits might lead to chaotic dynamics.

The theoretical developments of John Bush's group at MIT has been substantial \cite{MolBush13a,MolBush13b,ORB13}. For example, in an
investigation combining experiments and theory based on the model in
\cite{MolBush13b}, Wind-Willassen \emph{et al.} \cite{WMHB13} found, by numerical
integration of their model, that as $\gamma$ was increased, several
period-doubling sequences were generated, and finally that chaotic regions
appeared for sufficiently large parameter values. These outcomes turned out to
be in rather good agreement with their experimental results, which supported
the efficacy of the theoretical approach. There is also the rotating corral
study of Oza \emph{et al. }\cite{OHRB14} focused on the trajectory equation of
\eqref{e5.1} with a Coriolis force $F:=2m\boldsymbol{\omega}\times
\boldsymbol{\dot{r}}$. They found, using both analytical and numerical tools,
that as $Me$ increased, quantized stable circular orbits appeared, then tended
to wobble and destabilize, followed by period-doubling that seemed to lead to
chaotic regimes. In another related investigation, Oza \emph{et al.}
\cite{OWHRB14} studied the same trajectory equation for the horizontal
dynamics, with an eye toward behavior resulting from increasing memory.
Integrating the trajectory equation using the Adams--Bashforth method, they
found that as $Me$ was increased beyond the wobbling circular orbit phase, the
trajectories tended to drift into increasingly complicated periodic and
quasiperiodic trajectories and ultimately chaotic attractors as $Me$
approached unity. The fact that the conclusions of these last two papers is in
reasonably good agreement with the experimental results in Harris \emph{et al.
}\cite{HMFCB13} and Harris and Bush \cite{HarrisBush14}, lends considerable credence to
their fidelity.

More recently, Tambasco \emph{et al.} \cite{THORB} used and Adams--Bashforth
method to solve the trajectory equation of \eqref{e5.1} with a Coriolis,
Coulomb or simple harmonic potential force. They found that increasing
$\gamma$ (or $Me$) in the cases of Coriolis and Coulomb forces produced a
period-doubling cascade to chaos, while for a simple harmonic force the
transition to chaos resembled the mechanism described in Newhouse \emph{et
al.} \cite{NRT}. \ Kurianski \emph{et al}. \cite{KOB} also considered the
horizontal dynamics described by \eqref{e5.1} for a simple harmonic potential
force. They solved the corresponding trajectory equation using a combination
of an Adams--Bashforth integrator and Simpson's rule for the integral. The
solutions exhibited some fairly complicated periodic trajectories, including
trefoils and lemniscates, as well as quasiperiodic trajectories for fairly low
values of $\gamma$, which, as the forcing was increased, transitioned to
chaotic regimes that appeared to be decomposable into unstable quasiperiodic states.

Perrard and Labousse \cite{PL} studied the horizontal dynamics described by
the system \eqref{e5.1} confined by a radial horizontal force of the form
$F:=c\boldsymbol{r}$. By using an essentially spectral approach based on
expanding the wave field \eqref{e5.2} in a Fourier--Bessel series and
truncating it after a large number of terms, they reduced the system to a
finite-dimensional approximation comprising a system of ordinary differential
equations (ODEs) in the position, velocity and series coefficients. They then
integrated the system numerically and found the expected behavior; namely,
initially stable periodic orbits that destabilized and initiated a
period-doubling cascade to chaos with increasing $Me$. Moreover, the dynamical
behavior they found from their theoretical approach was in good agreement with
their experimental results.

Using a similar spectral-type approach, Budanur and Fleury \cite{BF} analyzed
the dynamics of a finite-dimensional approximation of the same trajectory
equation, modulo some symmetry reductions, studied in \cite{PL}. They
truncated the Fourier--Bessel expansion of the wave field to obtain a
$55$-dimensional system of ODEs, which they investigated numerically in
considerable detail. As a result, they found the expected period-doubling
route to chaos with increasing $Me$ mentioned above and a good deal more,
including symmetry breaking bifurcations, supercritical Neimark--Sacker
bifurcations \cite{Neimark, Sacker} of certain Poincar\'{e} sections generating invariant 2-tori,
supercritical Andronov--Hopf bifurcations, and what appear to be global
chaotic strange attractors associated with merging bifurcations related to
interactions of invariant submanifolds and stable manifolds of fixed points
and periodic orbits.  Similarly, Durey \cite{Durey2020}, observed Hopf bifurcations,
homoclinic bifurcations, and chaos in their Lorenz-like model.

There is also the quasi-potential model numerical of Galeano-Rios et al.
\cite{GMV17,GMV19} performed using a combination of finite differences
and a semi-implicit Euler method that produces dynamics that are in very good
agreement with the experimental and theoretical results in Wind-Willassen
\emph{et al}. \cite{WMHB13} and some of the other dynamical findings mentioned
above. They also investigated a new dynamical regime called superwalkers, which
was first observed by Valani \textit{et al.} \cite{VSS19}, wherein the
droplet bounces in place when the bath container is subject to a simple
harmonic vertical motion, but tends to walk exceptionally fast for mixed
harmonic oscillations.

In summary then, investigations of the existing continuous dynamical systems
models, conducted almost exclusively by numerical techniques, have revealed a
rich array of dynamics that apparently range from simple pitchfork
bifurcations to chaotic strange attractors.

\subsubsection{Dynamical results from discrete models}

Some very interesting dynamics have been found in the investigation of the
discrete dynamical system models described above. For example, Shirokoff
\cite{Shirokoff13} employed numerical simulation to study the dynamics of the
finite-dimensional difference equation \eqref{e5.8}, or equivalently the map
\eqref{e5.9} for two types of domains: $\Omega=\mathbb{R}^{2}$ or $\Omega=Q$,
a square region. He focused on the changes in the dynamics as parameters $F$
and $\mu$, representing the force of the oscillation and fluid viscosity,
respectively, were varied. The simulations showed that in the case when
$\Omega=\mathbb{R}^{2}$, increasing $F$ led to a pitchfork bifurcation
followed by a period-doubling cascade to chaos as the parameter was increased
and there was analogous dynamical behavior with respect to variation of $\mu$.
For the case of a square geometry, Shirokoff found, not surprisingly, that the
bifurcation behavior depended on the size of $Q$. For a large square, the
orbits tended toward a quasiperiodic attractor or a nearly dense attractor for
sufficiently large $F$. On the other hand, for a small square, a periodic or
quasiperiodic attractor emerged as the force increased, followed by apparently
chaotic regimes and ultimately very complicated attractors for large enough
$F$ values.

Simulation of the discrete dynamical system model of Durey and Milewski
\cite{DureyMilewski2017} also revealed some interesting dynamical behavior. Investigation of
their model for $\Omega=\mathbb{R}^{2}$ showed the existence of pitchfork
bifurcations and indications of period-doubling cascades with an increasing
force parameter, which we denote by $\mathcal{F}$ for simplicity. For
the same domain, but with the addition of a radial force, they observed
discrete quantization behavior in the form of separated invariant closed
curves that are initially stable and destabilize with increasing $\mathcal{F}$
while seemingly giving birth to stable tori and giving indications of a
transition to chaos. They also simulated their model for a pair of droplets,
which showed some of the bifurcations observed in the single droplet case and
several more very complex dynamic regimes.

\begin{figure}[hbtp	]
\centering
\stackinset{l}{4mm}{t}{0mm}{\textbf{\large (a)}}{\includegraphics[width = 0.32\textwidth]{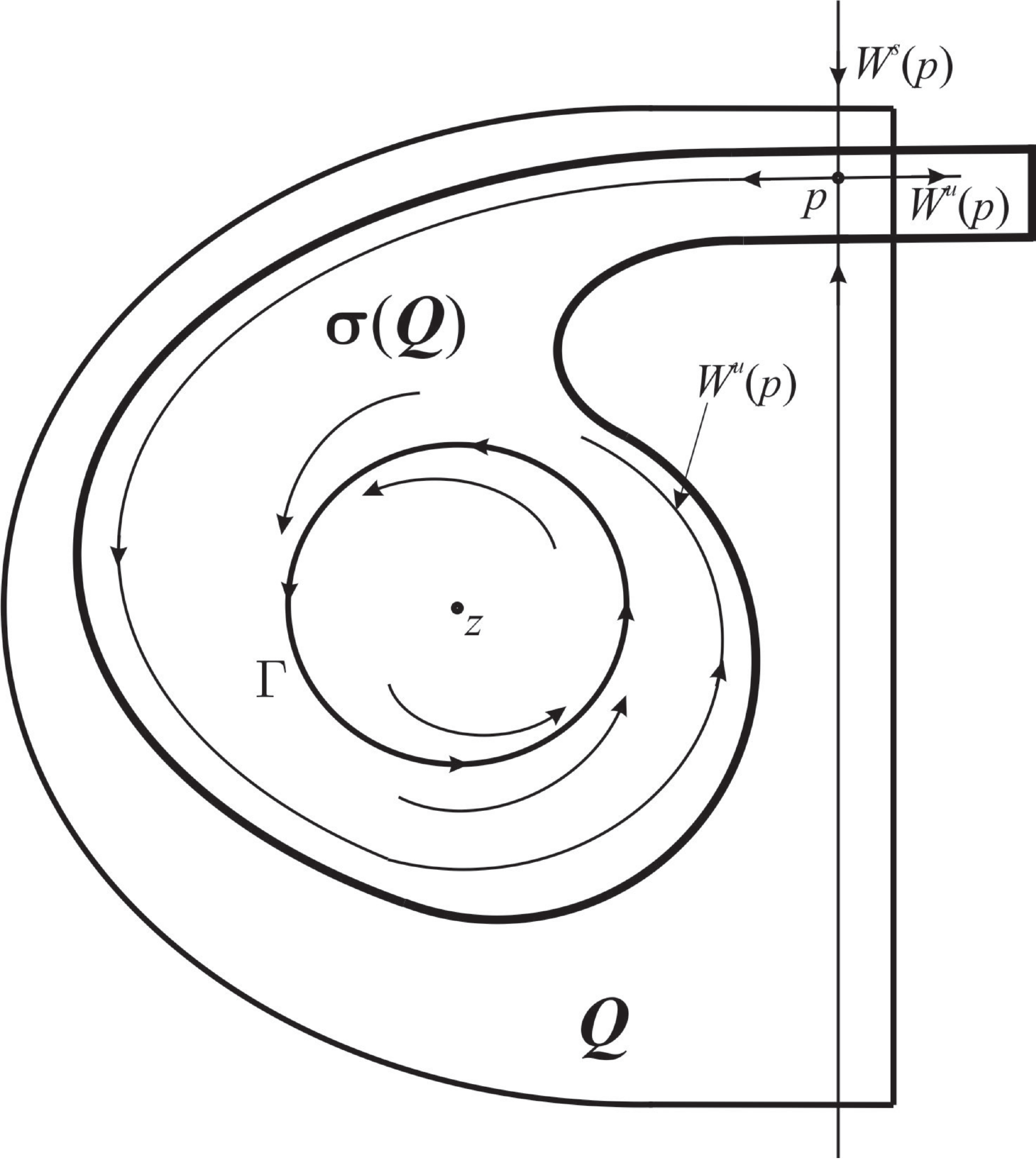}}
\stackinset{l}{4mm}{t}{0mm}{\textbf{\large (b)}}{\includegraphics[width = 0.32\textwidth]{S1}}
\stackinset{l}{4mm}{t}{0mm}{\textbf{\large (c)}}{\includegraphics[width = 0.32\textwidth]{S1}}
\stackinset{l}{4mm}{t}{2mm}{\textbf{\large (d)}}{\includegraphics[width = 0.32\textwidth]{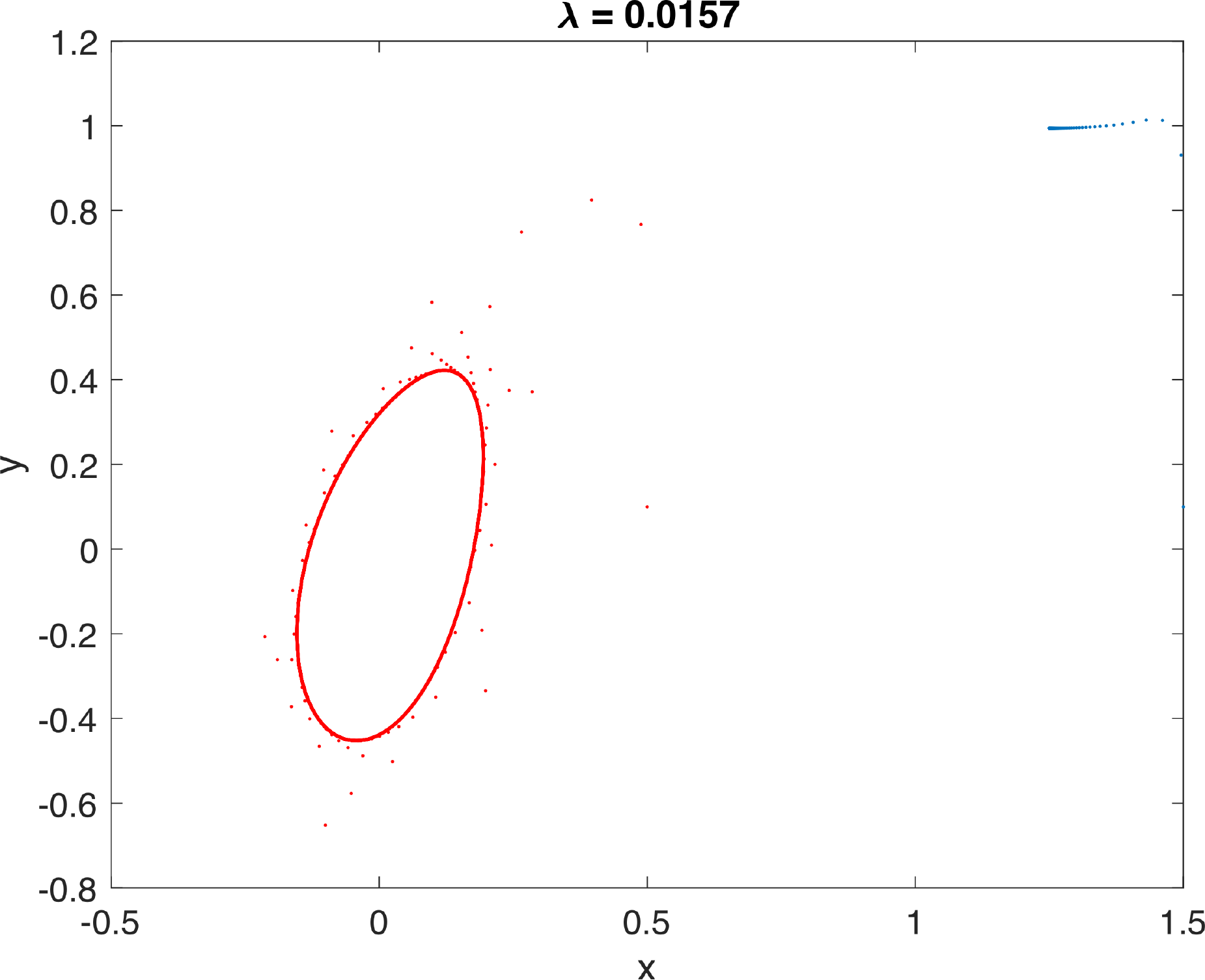}}
\stackinset{l}{4mm}{t}{2mm}{\textbf{\large (e)}}{\includegraphics[width = 0.32\textwidth]{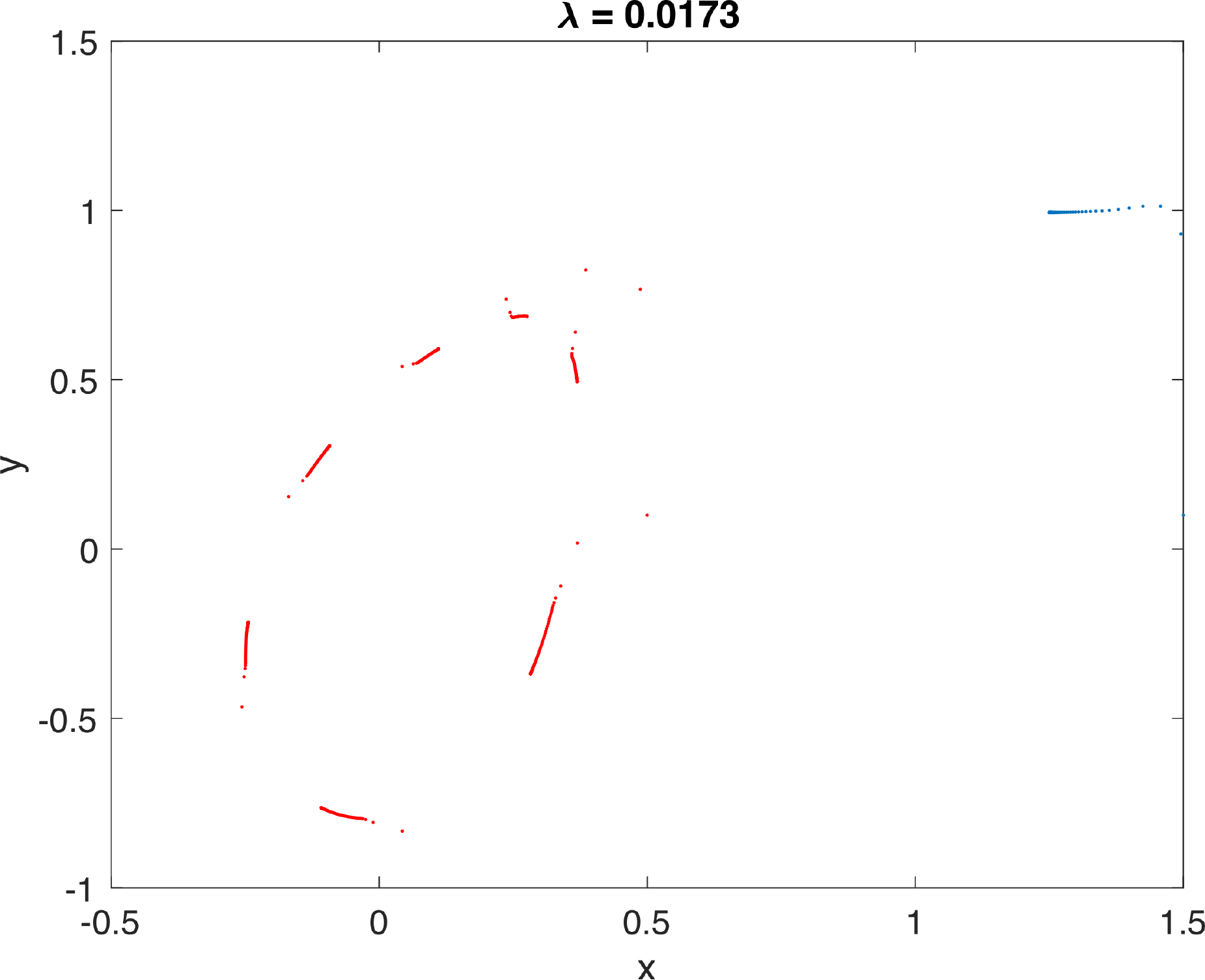}}
\stackinset{l}{4mm}{t}{2mm}{\textbf{\large (f)}}{\includegraphics[width = 0.32\textwidth]{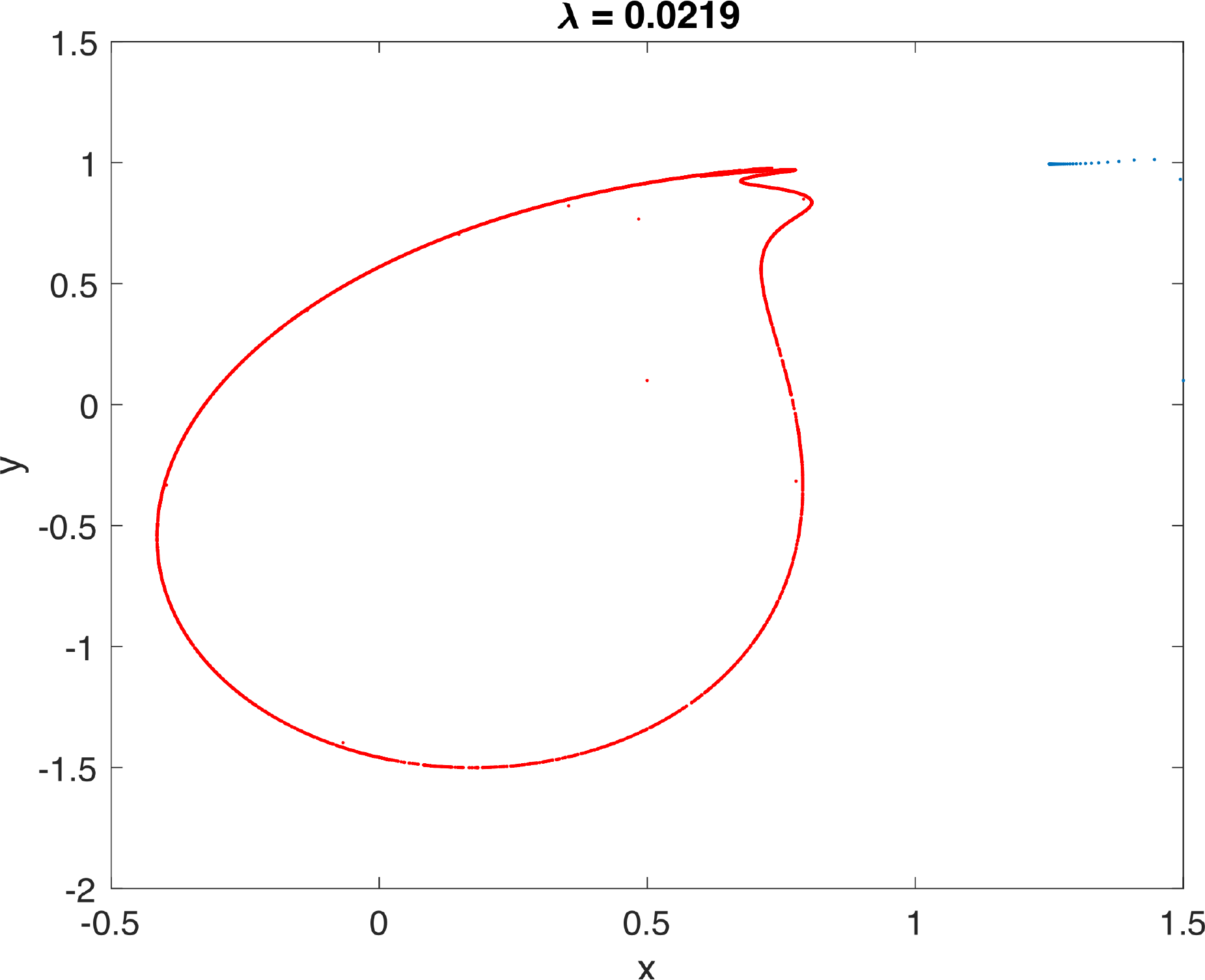}}
\caption{Illustration of the evolution of the $\sigma$-map through a homoclinic-type bifurcation.  Top row \textbf{(a)} - \textbf{(c)}:  diagrams of invariant manifolds as they evolve from left to right.  Bottom row \textbf{(d)} - \textbf{(f)}:  Simulations of the model under the bifurcation respectively corresponding to the diagrams \textbf{(a)} - \textbf{(c)}.  \textbf{(a)}, \textbf{(d)} Invariant manifolds are disjoint inducing regular dynamics.  \textbf{(b)}, \textbf{(e)} Once the invariant manifolds intersect tangentially, ``blinking'' behavior of the iterates begin; i.e., the iterates are not always dense within the invariant Jordan curve.  \textbf{(c)}, \textbf{(f)} The intersections are now transverse, and a stable chaotic strange attractor is generated.}\label{Fig: Sigma Map}
\end{figure}

In contradistinction to the state-of-the-art for continuous dynamical systems
models for walking droplet phenomena, there has been some rigorous analysis of
the dynamics. Simulation studies of \eqref{e5.12} indicated the existence of
supercritical Neimark--Sacker bifurcations \cite{Neimark, Sacker} and rather unusual transitions to
chaos (apparently due to the propinquity of growing attracting invariant
simple closed curves to stable manifolds of fixed points) with increasing $C$
and $\mu$, as well as indications of long-time behavior analogous to
quantization.  Gilet concluded that
the probability density function (PDF) for the model is proportional to
$1/\left\vert \Psi^{\prime}(x)\right\vert$. Most of these results were proved
or analytically verified in Rahman and Blackmore \cite{RahmanBlackmore16,RahmanBlackmore20}, where it was
proved that $G_{1}$ can have both supercritical and subcritical Neimark--Sacker
bifurcations \cite{Neimark, Sacker} with respect to either $C$ or $\mu$. Moreover, a rigorous
explanation was provided for the chaos inducing mechanism for $C$ sufficiently
close to unity, which involves the interaction of the basin of attraction of
an invariant closed curve attractor and the stable manifold of a hyperbolic
fixed point. This mechanism, which actually generates a chaotic strange
attractor, was abstracted in Rahman \emph{et al}. \cite{RJB17} in the form of a
paradigm called the $\sigma$-map (illustrated in Fig. \ref{Fig: Sigma Map}).

\begin{figure}[htbp]
\centering
\stackinset{r}{4mm}{t}{0mm}{\textbf{\large (a)}}{\includegraphics[width = 0.24\textwidth]{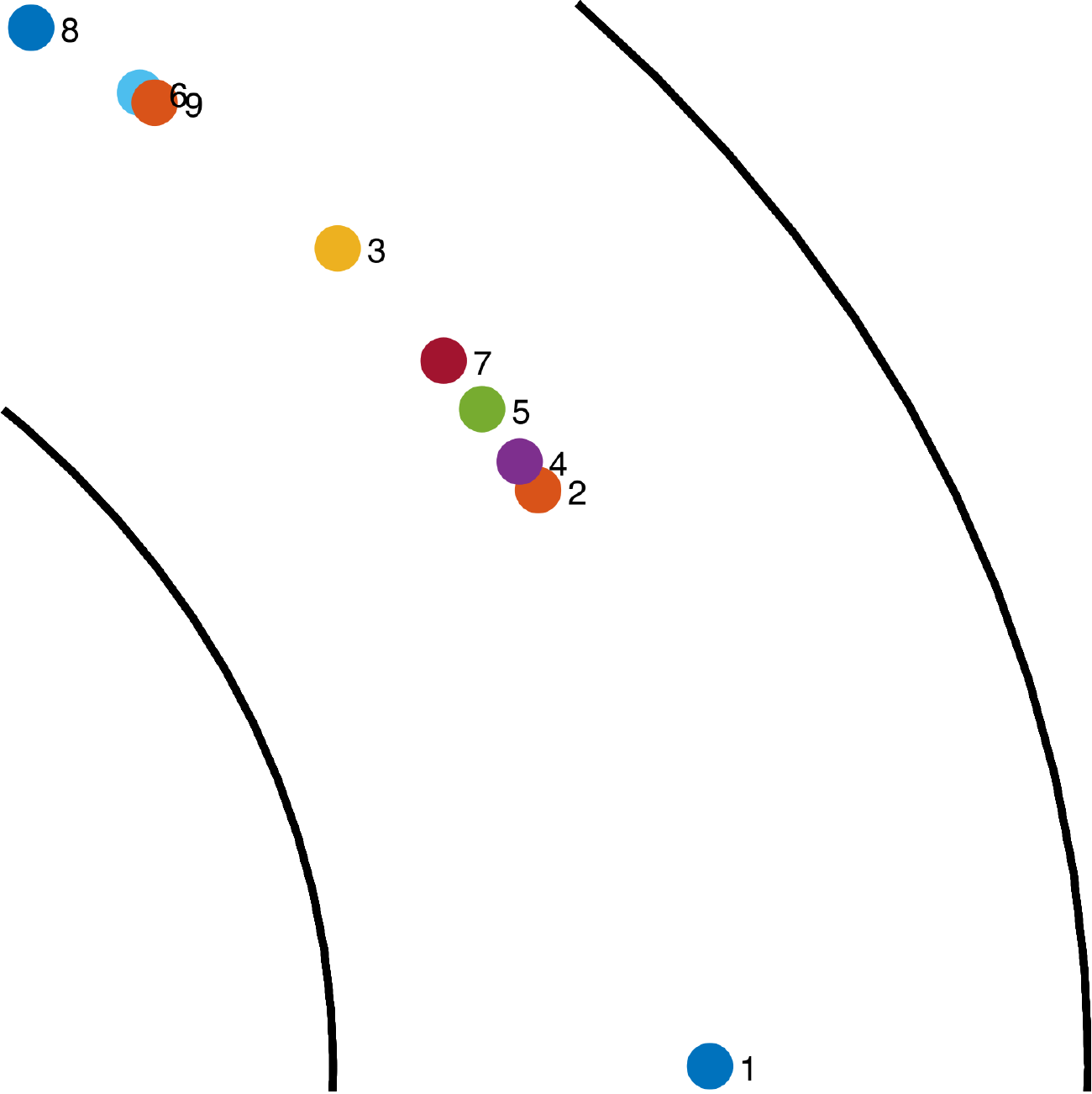}}
\stackinset{l}{4mm}{t}{1mm}{\textbf{\large (b)}}{\includegraphics[width = 0.32\textwidth]{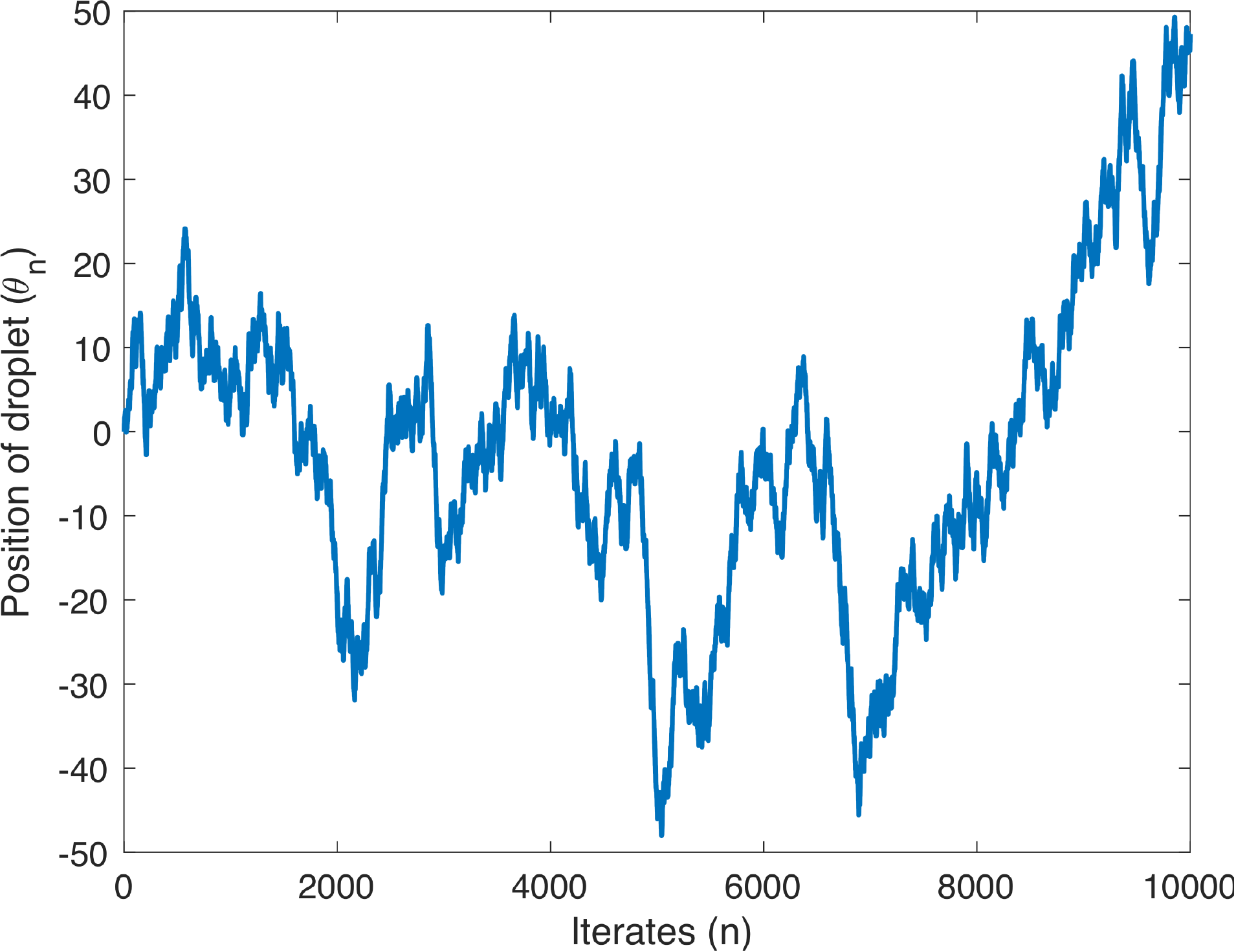}}
\stackinset{l}{4mm}{t}{0mm}{\textbf{\large (c)}}{\includegraphics[width = 0.4\textwidth]{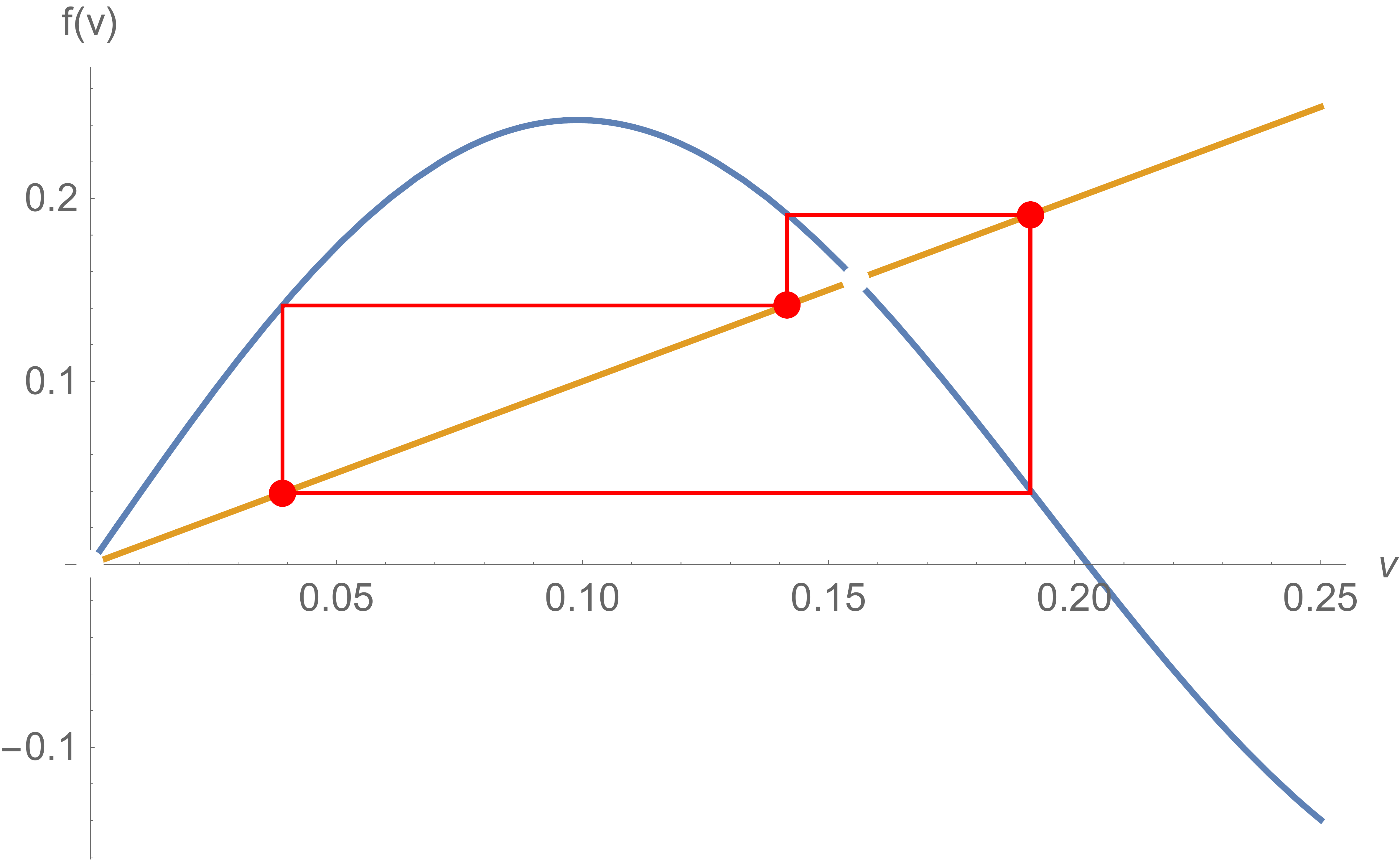}}
\caption{Chaos in the model of walkers on an annulus \eqref{Eq: SingleVelocity}\cite{Rahman18}.  \textbf{(a)}  Short time dynamics of the droplets showing sudden changes in velocity.  \textbf{(b)}  Long time dynamics showing seemingly random behavior.  \textbf{(c)}  Cobweb plot of a sample $3$-cycle in the map \eqref{Eq: SingleMap}.}\label{Fig:  StandardMapChaos}
\end{figure}

For the Standard map-like model of Rahman \cite{Rahman18}, they rigorously prove the existence of pitchfork and period doubling bifurcations and chaotic orbits as the damping parameter, $C$, is increased in \eqref{Eq: SingleMap}.  The two bifurcations are proved by conducting normal form calculations similar to those in Kuznetsov \cite{Kuznetsov}.  Through simulations of the model, one may observe chaos in the short timescale as the droplet suddenly changes velocity, as shown in Fig. \ref{Fig:  StandardMapChaos}, in a seemingly random fashion.  In the long timescale, the timeseries plot of Fig. \ref{Fig:  StandardMapChaos}, reveals complete disorder in the motion of the iterates.  To definitively prove the existence of chaotic iterates, they rigorously show the existence of a $3$-cycle (e.g. Fig. \ref{Fig:  StandardMapChaos}) and apply the theorem of Li and Yorke \cite{LiYorke75}.

The model given by \eqref{e5.13} assumes a single eigenmode, which is a
special case of the one Gilet \cite{Gilet16} considered for a circular corral that
incorporated twelve dominant Neumann eigenmodes, each with its own damping
factor $\mu_{k}\in\lbrack0,1]$. This leads to a 14-dimensional difference
equation model equivalent to the discrete dynamics of a map of the form
$\tilde{G}_{2}:\Omega\times\mathbb{R}^{12}\rightarrow\Omega\times
\mathbb{R}^{12}$ (which is essentially dynamically equivalent to $G_{2}$). The
simulation studies in \cite{Gilet16} showed that as the damping factors increases,
the trajectory map, which is $\tilde{G}_{2}$ projected onto $\Omega$, exhibits
stable periodic orbits that destabilize for sufficiently large $\mu_{k}$ via
Neimark--Sacker cycle bifurcations \cite{Neimark, Sacker}. Also observed was that a further increase
in the damping factors (with at least one of them slightly lower than unity)
induces chaotic regimes. Moreover, calculations of radial positioning
statistics for several simulations showed good agreement with corresponding
quantum (Schr\"{o}dinger equation) results, especially for the chaotic regimes.

\subsection{Some conjectures}

Based on what we know of the state-of-the-art of walking-droplet dynamics
research as outlined above and the current development of modern dynamical
systems theory, it is reasonable to posit the following conjectures:

\begin{itemize}
\item[(C1)] There is a class of smooth IDE systems including \eqref{e5.1} that
can be proved to exhibit pitchfork bifurcations, period-doubling cascades to
chaos and chaotic attractors generated by interactions of invariant manifolds
and possibly inertial manifolds containing global chaotic strange attractors.

\item[(C2)] It can be proved that the discrete dynamical system \eqref{e5.13}
exhibits analogs of all or most of the dynamical features in (C1). In
particular, a global chaotic strange attactor is created by the decreasing
distance between an expanding invariant torus and the stable manifold of a
fixed point as $C\rightarrow1$.

\item[(C3)] Informed by and building upon the work of Couder and Fort
\cite{CouderFort06}, Blanchette \cite{Blanchette2016}, Dubertrand \emph{et al. }\cite{DHSVBM},
Faria \cite{Faria2017} and Nachbin \emph{et al.} \cite{NMB} it should be possible
to construct an IDE model of walking droplet dynamics for a wide range of
topographies with two very useful features: It is amenable to rigorous
dynamical system analysis that reveals numerous important dynamical features
such as those in (C1) and (C2); and its dynamical predictions are in good
agreement with experiments and reliable simulations of other models.

\item[(C4)] The results in Gilet \cite{Gilet14,Gilet16} can be leveraged, mainly through
the choice of eigenmodes, to find discrete dynamical analogs of \eqref{e5.12}
and \eqref{e5.13} for variable topographies capable of producing
provable dynamics analogous to those in (C3), while also exhibiting reasonably
good agreement with experiments and simulations.
\end{itemize}

\section{Unsolved Problems}\label{Sec: Unsolved Problems}

In addition to the open problems associated with the conjectures (C1) - (C4)
above, we describe two more. One is the problem of developing a generalized
pilot-wave framework along the lines described in such investigations as Bush
\cite{Bush15a} and Fort and Couder \cite{FC} (see also Turton \emph{et al}.
\cite{TCB}).

The other is generalizing \eqref{e5.1} to an IDE for multiple droplets that
reliably predicts behavior and is amenable to dynamical systems analysis
capable of proving the existence of bifurcations and chaotic regimes and
possible chaotic strange attractors. In this effort, the pioneering work such
as that of Borghesi \emph{et al.} \cite{BMLEFC}, Oza \emph{et al.}
\cite{OSHMB} and Galeano-Rios \emph{et al.} \cite{G-RCCB} should prove very
helpful. Also, there is the related problem of extending \eqref{e5.12} and
\eqref{e5.13} to include multiple droplet dynamics.

Reduced Dynamical Systems models can also be used to significantly decrease computational expense of creating simulations.  For example, the models of Mol\'{a}\v{c}ek and Bush \cite{MolBush13a, MolBush13b} describe the physics of walking droplets in exquisite detail, which makes it computationally prohibitive to simulate the models long enough to observe the statistical behavior.  A reduced model, such as that of Oza \textit{et al.} \cite{ORB13}, is computationally fast enough to simulate on a laptop computer.

At the other end of the computational complexity spectrum, the simplified models of Gilet \cite{Gilet14, Gilet16}, Filoux \textit{et al.} \cite{FHV15}, and Rahman and Blackmore \cite{RJB17, Rahman18, RahmanBlackmore20}, have very fast computation times with fine-grained statistical data from the trajectories.  However, some of the physical accuracy has to be sacrificed, and while the models perform well for certain parameter regimes or a particular setup, they may miss some nuanced aspects of the phenomena.  Another challenge in this regard is to build models that retain their computational speed, but include enough of the physics details to precisely capture the statistical behavior shown in experiments.

\section*{Acknowledgments}

The authors would like to thank A. U. Oza, J. W. M. Bush, and D. Harris for fruitful discussions.  A special thank you to D. Harris for Figs. \ref{Fig: Experimental Setup} and \ref{Fig: Corral Experiment}.  A. R. appreciates the support of the Department of Applied Mathematics at UW, and D. B. appreciates the support of the Department of Mathematical Sciences at NJIT.

\bibliographystyle{ws-mplb}
\bibliography{Bouncing_droplets}

\end{document}